\documentclass[]{elsart}
\usepackage{graphics}
\usepackage{graphicx}
\usepackage{epsfig}
\usepackage{amssymb}
\usepackage{psfrag}
\usepackage{dcolumn}
\usepackage{natbib}

\journal{Int. J. Heat and Fluid Flow}
\begin{document}


\begin{frontmatter}

\title{Wall-oscillation conditions for drag reduction in turbulent channel flow}
\author[imp]{Pierre Ricco\corauthref{cor}},
 \ead{pierre.ricco@kcl.ac.uk}
 \author[pdm]{Maurizio Quadrio}
\corauth[cor]{Corresponding author. Present address: Department of Mechanical Engineering, King's College London, Strand, London WC2R 2LS, United Kingdom.}

\address[imp]{Department of Mathematics, Imperial College London\\
180 Queen's Gate - London SW7 2BZ, United Kingdom}

\address[pdm]{Dipartimento di Ingegneria Aerospaziale, Politecnico di Milano,\\
via La Masa, 34 - 20156 Milano, Italy}

\end{frontmatter}

\newpage

\begin{frontmatter}
\begin{abstract}
The drag reduction properties of a turbulent channel flow modified by spanwise sinusoidal oscillations of the walls are investigated by direct numerical simulations. The work is based on the linear relation between the drag reduction and the parameter $S$, function of the maximum wall velocity and the period of the oscillation. This quantity, first determined by \citet{choi-xu-sung-2002} and later studied by \citet{quadrio-ricco-2004b}, has been found through physical arguments pertaining to the action of the oscillating Stokes layer on the near-wall turbulence dynamics. The predictive potential of the scaling parameter is exploited to gain insight into the drag-reducing effects of the oscillating wall technique. The period of oscillation which guarantees the maximum drag reduction for a given maximum wall displacement is studied for the first time. The issue of the minimum intensity of wall forcing required to produce a non-zero drag reduction effect and the dependence of the drag reduction on the Reynolds number are also addressed. The drag reduction data available in the literature are compared with the prediction given by the scaling parameter, thus attaining a comprehensive view of the state of the art.
\end{abstract}

\end{frontmatter}

\newpage

\section{Introduction}

This paper presents a numerical investigation of a turbulent channel flow with sinusoidal spanwise oscillations of the walls. The flow over the oscillating walls results from the combination of two simpler flows, i.e. a canonical turbulent channel flow in the streamwise direction and an oscillating boundary-layer-type flow in the transversal (spanwise) direction. The most relevant characteristic of this modified turbulent flow is the time-sustained reduction of the streamwise wall-shear stress, first pointed out by \citet{jung-mangiavacchi-akhavan-1992}. The drag reduction effect is caused by the weakening of the relevant turbulence-producing events in the vicinity of the wall \citep{akhavan-etal-1993,dhanak-si-1999,choi-clayton-2001,dicicca-etal-2002,iuso-etal-2003,ricco-2004,xu-huang-2005,zhou-ball-2006}, but the precise details are still poorly understood. It has also been established \citep{baron-quadrio-1996,quadrio-ricco-2004b} that a net energetic saving of the order of 10\% (determined by taking into account the power spent to move the walls against the viscous resistance of the fluid) can be obtained by carefully tuning the parameters of the oscillation. Another property of this flow that will be exploited in the following is that, once the flow field is averaged along the streamwise and spanwise homogeneous directions, the spanwise velocity profile agrees with the laminar solution of the so-called second Stokes problem \citep{quadrio-sibilla-2000,choi-xu-sung-2002,quadrio-ricco-2003}, hence uncoupling from the complex dynamics of the all-encompassing turbulence. In the spirit of the wall-oscillation technique, research works have appeared on the drag reduction effects of forcing the turbulence by spanwise travelling waves \citep{du-karniadakis-2000,du-symeonidis-karniadakis-2002,karniadakis-choi-2003,zhao-wu-luo-2004,itoh-etal-2006,yoon-etal-2006}, by spanwise oscillating Lorentz forces \citep{berger-etal-2000,pang-choi-2004,breuer-park-henoch-2004,lee-sung-2005}, by spanwise oscillating suction and blowing \citep{segawa-etal-2005}, and by steady streamwise oscillations of the spanwise wall velocity \citep{quadrio-viotti-luchini-2007}. The wall oscillation has also been shown to be effective in reducing the growth rate of the most unstable G$\ddot{\mbox{o}}$rtler vortex developing on a concave surface \citep{galionis-hall-2005}.

An important step towards practical applications of the oscillating-wall technique is the recent finding of a scaling parameter $S$ that is suggested to be related to the amount of drag reduction. This parameter depends on the quantities defining the sinusoidal oscillation, namely the period of oscillation $T$ and the maximum wall velocity $W_m$. (A third parameter describing the oscillation is the peak-to-peak wall displacement $D_m$, which is $D_m = W_m T/\pi$ for a sinusoidal waveform.) \citet{choi-xu-sung-2002} have correlated their drag reduction direct numerical simulation (DNS) data with $S$, which was found from physical arguments pertaining to the interaction between the spanwise laminar Stokes layer and the near-wall turbulence. A least-squares fit yielded a power-law expression for $S$. \citet{quadrio-ricco-2004b} (denoted by QR hereafter) have recently improved the analysis by \citet{choi-xu-sung-2002} on the basis of a DNS dataset of a turbulent channel flow modified by the motion of both walls, for a Reynolds number $Re_\tau=200$ defined by the friction velocity of the natural turbulent flow and half the distance between the channel walls. The accuracy of their dataset brought to light a linear relation between the drag reduction and $S$, which was suggested to hold as long as $T$ remains smaller than a typical integral time scale of the wall turbulence.

The main objective of the present work is to investigate further the possibility of estimating the drag reduction through the parameter $S$. In QR the correlation analysis between $S$ and the drag reduction was carried out only with their DNS data. In the present analysis all the drag reduction datasets available in the literature are employed to compute a new, and more general, least-squares fit. A useful by-product of this procedure is that each dataset (both from experiments and DNS) can be discussed in terms of its difference from the fit. Since the literature data show a considerable scatter, comparing the actual measurements of drag reduction with the predictions via $S$ permits an indirect assessment of their accuracy.

The linear fit allows us to identify and study easily two distinct periods of wall oscillation that are optimal in terms of drag reduction. Although in DNS studies the oscillation parameters can be chosen freely, $T_{opt,W}$, the optimal period of oscillation for fixed maximum wall velocity, has usually been the quantity of interest. On the other hand, an experimental campaign is likely to reveal $T_{opt,D}$, the optimal period at fixed maximum wall displacement. Indeed, in a laboratory the wall motion is usually produced by a crank-slider mechanism, which allows varying $T$ for fixed $D_m$. QR however noted that $T_{opt,D}$ had never been observed by experimentalists, probably because the flow at these high frequencies of oscillation is difficult to test. The difference between $T_{opt,D}$ and $T_{opt,W}$ was left unnoticed until QR pointed out that $T_{opt,D}$ should be smaller than $T_{opt,W}$, which is known to be constant at $T_{opt,W}^+ \approx 125$ (when scaled by the friction velocity of the undisturbed flow, Karniadakis and Choi, 2003). QR's analysis also showed that $T_{opt,D}$ should be a function of $D_m$ itself. Although their limited data appeared to validate these unconfirmed conjectures, it is an important aim of the present study to verify the existence of $T_{opt,D}$ and to assess its properties.

The analytical expression for $S$ yields additional information. As a further aim, we shall use $S$ to seek the smallest values of the oscillation parameters which guarantee a non-zero drag reduction effect. These oscillating conditions are referred to as ``minimal" throughout the paper. The same analysis will also allow us to determine the oscillatory conditions needed to attain a specified amount of drag reduction and of net energy saving.

We shall also address the important issue of the effect of Reynolds number on the drag reduction. This point is still open to discussion, as the conclusions of previous studies do not fully agree. The experimental analysis of a turbulent boundary layer by \citet{ricco-wu-2004} has shown that increasing the Reynolds number based on the momentum thickness and free-stream velocity from $Re_\theta$=500 to 1400 (with $D_m^+ \approx 240$) had no influence on the drag reduction.
\citet{choi-graham-1998}'s experimental results in a pipe flow modified by circular wall oscillations have also indicated that changing the Reynolds number based on the bulk velocity and the pipe diameter from $Re$=23,300 to 36,300 gives variations of drag reduction which are of the order of the uncertainty range. The DNS of a turbulent channel flow conducted by \citet{choi-xu-sung-2002} has instead revealed that the reduction in wall friction can be halved by increasing the Reynolds number from $Re_\tau=100$ to $Re_\tau=400$. We shall carry out a few DNS to ascertain whether or not increasing $Re_\tau$ affects the drag reduction.

The paper is organized as follows. In Section~\ref{sec:procedure}, the numerical procedure and the discretization parameters are described. The main steps of the analysis by QR are briefly recalled in Section~\ref{sec:laminar-analysis}, where the parameter $S$ and the function $T_{opt,D}$=$T_{opt,D}(D_m)$ are defined. Section~\ref{sec:Topt} discusses the quality of the analytical prediction of $T_{opt,D}$ on the basis of the DNS results. The minimal wall forcing conditions and the dependence of the drag reduction on the oscillatory parameters are presented in Section~\ref{sec:minimal}. Section~\ref{sec:Reynolds} describes the effect of the Reynolds number on the drag reduction. Section~\ref{sec:validation-datasets} presents the drag reduction data available in the literature, providing an evaluative overview on the state of the art in terms of a comparison between the amounts of drag reduction and the estimates based on $S$. Section~\ref{sec:conclusions} is devoted to a summary.

\section{Numerical procedure}
\label{sec:procedure}

\begin{figure}
\vspace{1cm}
\centering
\psfrag{F}{Mean flow}
\psfrag{x}{$x$}
\psfrag{y}{$y$}
\psfrag{z}{$z$}
\psfrag{Lx}{$L_x$}
\psfrag{Ly}{$L_y$}
\psfrag{Lz}{$L_z$}
\psfrag{W}{$W=W_m \sin \left(\frac {2\pi}{T} t \right)$}
\centering
\includegraphics[width=\textwidth]{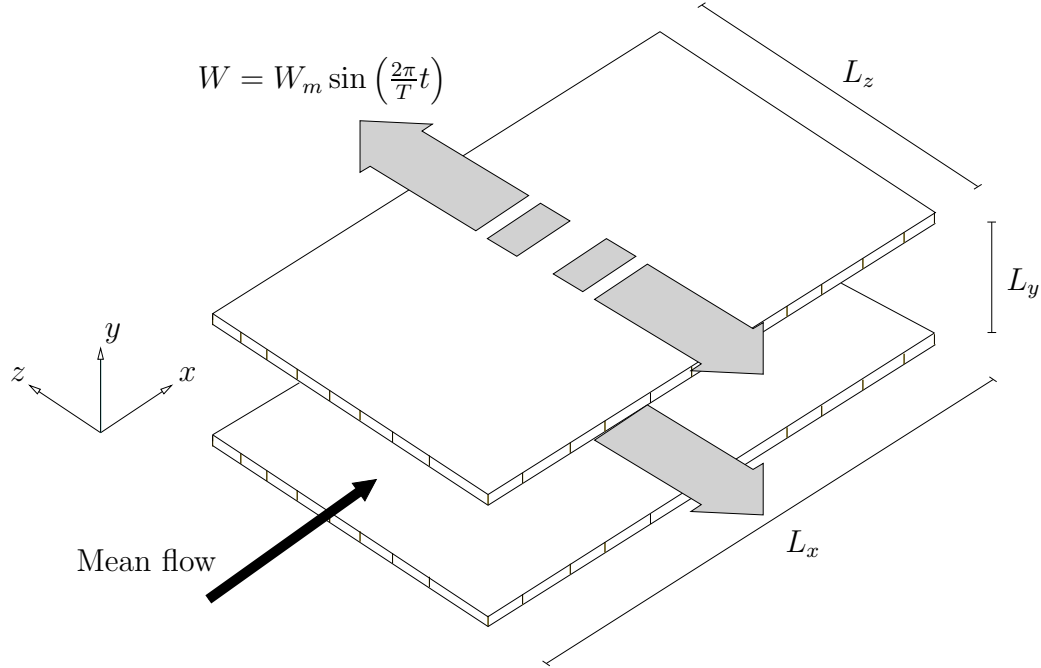}
\caption{Schematic of the physical domain.}
\label{fig:figure_1}
\end{figure}

We have studied the turbulent flow in a channel with moving walls through direct numerical simulations of the incompressible Navier--Stokes equations. The walls move in phase along the spanwise direction with velocity
\[
W(t;W_m,T) = W_m \sin \left( \frac{2 \pi t}{T} \right).
\]
Figure \ref{fig:figure_1} displays a sketch of the computational domain.

Our pseudo-spectral solver is described in \citet{luchini-quadrio-2006}: it is based on Fourier expansions in the homogeneous directions and on fourth-order accurate, compact finite-difference schemes for the discretization of the differential operators in the wall-normal direction. Aliasing errors in the computation of the non-linear terms are eliminated by expanding the flow variables into a (at least) $3/2$ larger number of modes for each homogeneous direction before transforming from the Fourier to the physical space. The temporal integration is carried out with a partially-implicit procedure: a third-order, low-storage Runge-Kutta method for the convective terms, and a second-order Crank-Nicolson scheme for the viscous terms. The mixed spatial discretization is advantageous from the viewpoint of parallel computing \citep{luchini-quadrio-2006}, and allows employing computing machines connected by standard networking hardware to achieve a large computational throughput.

The simulations described in this paper have been carried out on a computing system available in dedicated mode at the University of Salerno, made by 64 machines, each with two Opteron CPUs. We have performed the calculations at two values of the Reynolds number: $Re_\tau=200$, 400 based on $h$, half the distance between the channel walls, and on $u_\tau$, the friction velocity of the uncontrolled case. The computational parameters for $Re_\tau=200$ have been chosen to replicate those in QR, if exception is made for a slightly finer wall-normal discretization. We have employed a computational box with the following dimensions: $L_y = 2 h$, $L_x = 21 h$, $L_z = 4.2 h$ in the wall-normal, streamwise and spanwise directions, respectively. We have used 160 grid points in the wall-normal direction, and $321$ and $129$ Fourier modes in the streamwise and spanwise directions. For the $Re_\tau=400$ case, we have employed 256 grid points in the wall-normal direction, and $385$ and $321$ Fourier modes in the streamwise and spanwise directions. Each simulation has been run for 40,000 time steps with $\Delta t^+ = 0.2$. Throughout the paper, the + superscript indicates quantities scaled by the inner units of the unperturbed turbulent flow. A few calculations with very high $W_m$ have required an even smaller (up to one half) time step, due to stability constraints. A single case at $Re_\tau=200$ has taken 11 seconds for one time step and 5 days for the full simulation with 8 machines. The wall-clock time for the whole set of simulations has been slightly more than two weeks with the full use of the computational power.

For the $Re_\tau=200$ case, we have carried out one simulation for the canonical flow and 28 simulations for different $(T,W_m)$ pairs. The skin-friction coefficient for the fixed-wall case is $C_f = 2 \tau_x / (\rho U_b^2) = 7.94 \times 10^{-3}$, where $\tau_x$ is the time- and space-averaged streamwise wall-shear stress, $\rho$ is the fluid density and $U_b$ is the bulk velocity. This value essentially coincides with the value estimated by the following empirical formula given in \citet{pope-2000} at page 279:
\begin{equation}
\label{eq:cf-re-tau}
C_f = 0.0336 Re_\tau^{-0.273}.
\end{equation}
Three values of $D_m$ have been considered, $D_m^+ = 100, 200, 300$. At a given $D_m$, we have explored the existence of $T_{opt,D}$ by varying $T$ along the hyperbola $W_m=D_m \pi/T$ in the $(T,W_m)$ space. The amount of drag reduction has been determined by a procedure (QR) which involves discarding the initial temporal transient, averaging over time and along the homogeneous directions, and considering both walls to enlarge the statistical sample. For the $Re_\tau=400$ case, we have carried out one simulation for the canonical flow ($C_f = 6.495 \times 10^{-3}$, only 1\% smaller than the value given by (\ref{eq:cf-re-tau})) and three simulations with $W_m^+$=12 and $T^+$=30, 125 and 200.

\section{Laminar analysis}
\label{sec:laminar-analysis}

\begin{figure}
\psfrag{R}{$R_{vw}$}
\psfrag{y}{$y^+$}
\centering
\includegraphics[width=\textwidth]{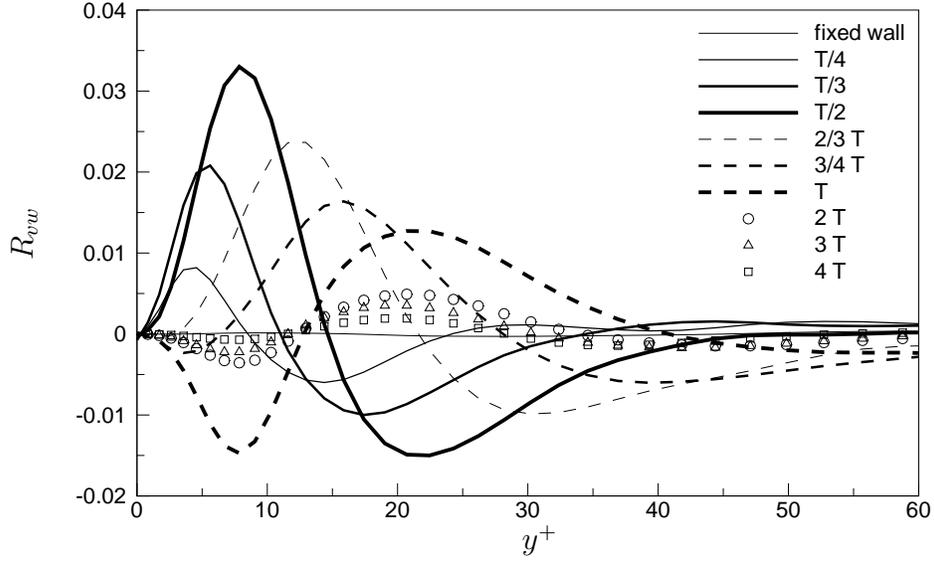}
\caption{$R_{vw}=\partial \overline{v' w'}^+/\partial y^+$ as a function of $y^+$ at different oscillation phases from the start-up of the wall motion ($W_m^+=18$, $T^+=125$).}
\label{fig:figure_2}
\end{figure}

The analysis is based on the close agreement between the space-averaged (along the homogeneous directions) spanwise flow and the laminar solution for the second Stokes problem, defined by $w^+_s \left( y^+,t^+;W_m^+,T^+ \right)$ \citep{quadrio-sibilla-2000,choi-xu-sung-2002,quadrio-ricco-2003}. At relatively high frequencies, the space-averaged spanwise flow uncouples from the space-averaged streamwise flow, although the spanwise turbulent fluctuations are significantly altered  by the wall motion \citep{quadrio-ricco-2003}. We can explain this agreement as follows. The space-averaged spanwise momentum equation is
\[
\frac{\partial \overline{w}^+}{\partial t^+}\Bigg{|}_{y^+} = \frac{\partial^2 \overline{w}^+}{\partial y^{+ 2}}\Bigg{|}_{t^+} - R_{vw},
\]
where $R_{vw}=\partial \overline{v' w'}^+/\partial y^+|_{t^+}$. The barred quantities are averaged along $x$ and $z$, $w$ is the turbulent spanwise velocity, and $v'$ and $w'$ are the wall-normal and spanwise velocity components fluctuating about their corresponding space-averaged quantities. The equation describing the laminar Stokes flow is obtained by replacing $\overline{w}$ with $w_s$, and by setting $R_{vw}=0$.
Fig. \ref{fig:figure_2} shows $R_{vw}$ at different oscillation phases from the start-up of the wall motion for $W_m^+=18$ and $T^+=125$. It initially increases, reaches a maximum when $t \approx T/2$, and then decreases as the transient elapses after a few oscillation periods. As $R_{vw}$ is negligible once the new modified turbulent flow is established, it follows that $\overline{w}^+ \approx w^+_s$ as $t^+ \rightarrow \infty$.
The fact that $\overline{w}^+$ does not coincide with $w^+_s$ during the initial transient is reflected in the time history of the space-averaged spanwise wall-shear stress $\tau_z$. It is observed in Fig. 8 at page 12 in \citet{quadrio-ricco-2003} that this quantity deviates slightly from the laminar value at $t \approx T/2$, when $R_{vw}$ reaches its maximum value.

The scaling parameter $S$ is now introduced. It has been expressed by \citet{choi-xu-sung-2002} as the product of $a_m^+$, the maximum acceleration of the Stokes layer at a specified wall-normal location, and $\ell_w^+$, a distance from the wall at which the oscillating layer directly affects the turbulent flow.

At the root of this procedure lies the laminar solution for the second Stokes problem \citep{batchelor-1967}:
\begin{equation}
\label{eq:stokes}
w^+_s \left( y^+,t^+;W_m^+,T^+ \right) = W_m^+  \exp \left( - y^+ \sqrt{\pi / T^+} \right)
\sin \left( \frac{2 \pi}{T^+} t^+ - y^+ \sqrt{\frac{\pi}{T^+}} \right).
\end{equation}
The maximum spanwise acceleration $a_m^+$ at a wall-normal distance $y^+=\ell_a^+$ is obtained by differentiating (\ref{eq:stokes}) with respect to $t^+$:
\[
a_m^+ = \frac{2 \pi W_m^+}{T^+} \exp \left(- \ell_a^+ \sqrt{\pi / T^+}\right).
\]
It is further required that the maximum spanwise velocity of the Stokes layer at $y^+=\ell_w^+$ be larger than a typical value $W_{th}^+$ of the spanwise velocity fluctuations. The wall forcing must be intense enough for the Stokes layer to influence the turbulent fluctuations and disrupt the near-wall turbulence-producing cycle. By imposing the above condition and by using (\ref{eq:stokes}), it follows
\[
\ell_w^+ = \sqrt{\frac{T^+}{\pi}} \ln \left(\frac{W_m^+}{W_{th}^+}\right).
\]
The quantities $a_m^+$ and $\ell_w^+$ are united into the expression for $S$:
\begin{equation}
\label{eq:S}
S = \frac{a_m^+ \ell_w^+}{W_m^+} = 2 \sqrt{\frac{\pi}{T^+}}
\ln \left( \frac{W_m^+}{W_{th}^+} \right) \exp \left(- \ell_a^+ \sqrt{\pi / T^+}\right).
\end{equation}
$\ell_a^+$ and $W_{th}^+$ have been determined by maximizing the correlation coefficient $C_S$ between the drag reduction data and $S$. In QR, $C_S=0.99$, $\ell_a^+=6.3$ and $W_{th}^+=1.2$ (which remarkably agrees with the maximum r.m.s. of the turbulent spanwise velocity fluctuations, $w^+_{rms} \approx 1.1$ \citep{kim-moin-moser-1987}) and
\begin{equation}
\label{eq:fit}
DR(\%) = S_1 S + S_2,
\end{equation}
where $DR(\%)$ is the drag reduction, $S_1$=130.6 and $S_2$=-2.7.

An expression for $T^+_{opt,D}$, the period of oscillation which guarantees the maximum drag reduction for fixed $D_m$, is found by setting:
\[
\frac{\partial DR}{\partial T^+}\Bigg{|}_{D_m^+} = S_1 \frac{\partial S}{\partial T^+}\Bigg{|}_{D_m^+} = 0, \hspace{0.25 in}
\frac{\partial^2 DR}{\partial T^{+ 2}}\Bigg{|}_{D_m^+} = S_1 \frac{\partial^2 S}{\partial T^{+ 2}}\Bigg{|}_{D_m^+} < 0.
\]
The second condition has been verified graphically via (\ref{eq:S}). After eliminating $W_m^+ = \pi D_m^+ / T^+$ in (\ref{eq:S}), it follows:
\begin{equation}
\label{eq:Topt-Dm}
\left( \ell_a^+\sqrt{\frac{\pi}{T^+_{opt,D}}}-1 \right) \ln\left( \frac{\pi D_m^+}{T^+_{opt,D} W_{th}^+} \right) = 2.
\end{equation}
Differently from $T^+_{opt,W}$, which does not depend on $W_m^+$, $T^+_{opt,D}$ increases monotonically with $D_m^+$ and is smaller than $T^+_{opt,W}$. The latter is obtained by $\partial S / \partial T^+ \big{|}_{W_m^+}= 0$, i.e. $T^+_{opt,W}=\pi (\ell_a^+)^2 \approx 125$, which agrees with the values in \citet{jung-mangiavacchi-akhavan-1992}, \citet{dhanak-si-1999}, \citet{quadrio-sibilla-2000} and QR.

QR have also shown that $S$ scales linearly with the drag reduction only for $T^+ \leq 150$. Fortunately, it occurs that $T^+_{opt,D}, T^+_{opt,W} <$150, so that the prediction of the optimal periods based on the linear relation between the drag reduction and $S$ is valid. These quantities are not well correlated when the oscillation uncouples from the near-wall turbulence dynamics, namely when half the period of oscillation is larger than a typical pseudo-Lagrangian time scale representing the survival time of the longest-lived structures ($\approx 60$ time units) \citep{quadrio-luchini-2003}. When $T$ is large, the near-wall structures have enough time to develop their inner dynamics between successive sweeps of the Stokes layer.
In this limit, the flow adapts to a new quasi-steady three-dimensional condition, where time can be treated as a parameter and the drag-reducing effect of the Stokes layer is lost.

\section{Drag reduction scaling and optimum period at fixed $D_m$}
\label{sec:Topt}

\begin{figure}
\psfrag{D}{$DR (\%)$}
\psfrag{S}{$S$}
\centering
\includegraphics[width=\textwidth]{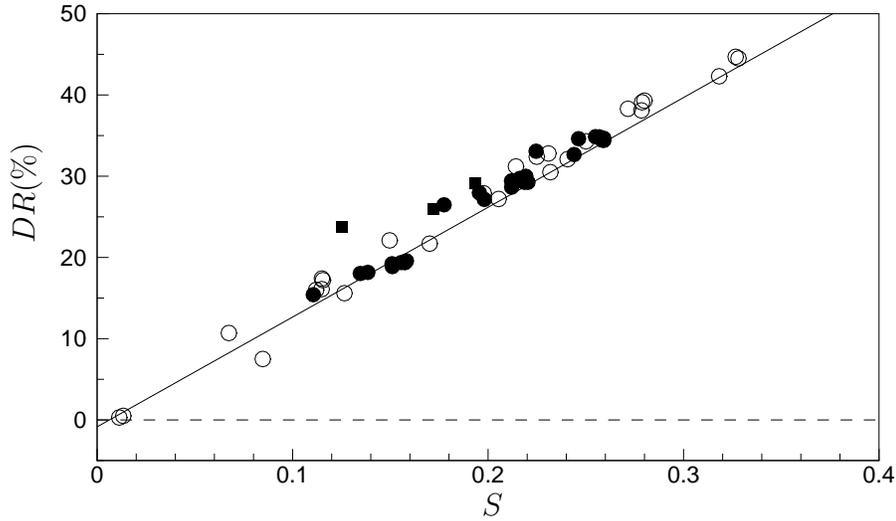}
\caption{Drag reduction data from the present study ($\bullet$ and $\blacksquare$) and from QR ($\circ$) as function of the scaling parameter $S$. Refer to Section~\ref{sec:Topt} for discussion on the data indicated by black squares.}
\label{fig:figure_3}
\end{figure}

Fig. \ref{fig:figure_3} shows the drag reduction DNS dataset versus $S$ for $Re_\tau=200$ produced for the present analysis, together with the data by QR for $T^+ \leq 150$. The oscillating conditions are largely different, but the data collapse well on the straight line. The linear regression (\ref{eq:fit}) has been recomputed by grouping our DNS data with most of the data available in the literature, which are analysed in Section~\ref{sec:validation-datasets}. The only discarded datasets were the ones by \cite{choi-xu-sung-2002} at $Re_\tau=100$ because during the oscillation the Reynolds number was too low and the ones by \cite{dhanak-si-1999}, whose analysis was based on a simplified model and not on the Navier-Stokes equations. The new correlation parameters are $C_S=0.92$, $S_1$=135.11, $S_2$=-0.85, $\ell_a^+=6.2$ and $W_{th}^+=1.7$, which are essentially unchanged from the analysis in QR for $Re_\tau=200$. These new values have been used for the present data analysis.

\begin{figure}
\psfrag{T}{$T^+$}
\psfrag{D}{$DR (\%)$}
\centering
\includegraphics[width=\textwidth]{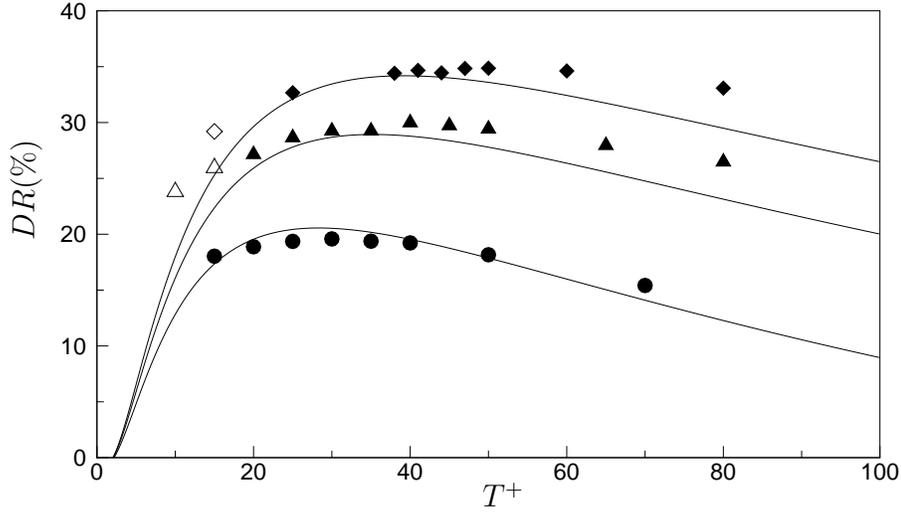}
\caption{Drag reduction data as function of period of oscillation at fixed $D_m$, for $D_m^+=100$ ($\bullet$), 200 ($\blacktriangle$ and $\triangle$) and 300 ($\blacklozenge$ and $\diamond$). Solid lines represent the corresponding estimates based on (\ref{eq:S}) and (\ref{eq:fit}). Refer to Section~\ref{sec:Topt} for discussion on the data indicated by the open symbols.}
\label{fig:figure_4}
\end{figure}

The results of the present simulations, designed to identify $T_{op t,D}$, are compared in Fig. \ref{fig:figure_4} with the prediction obtained by (\ref{eq:S}) and (\ref{eq:fit}). Good agreement occurs except for three data points at low $T$ (open symbols in Fig. \ref{fig:figure_4} and black squares in Fig. \ref{fig:figure_3}). For these points, the predicted values are lower than the actual DNS data. We have not been able to explain this behaviour at very small $T$, which corresponds -- being the displacement fixed -- to very large $W_m$ ($W_m^+>$40). We have first verified that the space-averaged spanwise velocity profile at various phases still agrees with the laminar solution, from which (\ref{eq:S}) is determined. As a further check, the Reynolds number $Re_\delta$ for the spanwise oscillating flow, based on $W_m$ and on the Stokes layer thickness $\delta_s = \sqrt{\nu T/\pi}$ (where $\nu$ is the kinematic viscosity of the fluid) is compared with the critical Reynolds numbers for stability or transition of the Stokes flow, although this analysis is obviously not rigorous in that we consider a space-averaged profile and not a purely laminar flow. Despite the wall velocity being as high as twice the centreline velocity, $Re_\delta = W_m^+\sqrt{T^+/\pi}$ is not high, since $T$ is low. For our data, $Re_\delta= 80-140$, which is lower than $\approx 1400$ found by \citet{blennerhassett-bassom-2006} as the lowest critical Reynolds number for an oscillatory boundary layer between two parallel plates, and $\approx 550$ reported by \citet{vittori-verzicco-1998} as their transition threshold.

However, these discrepancies do not limit the possibility of employing $S$ to study successfully the flow because all the relevant drag reduction features occurs for smaller $W_m^+$. Indeed, (i) $T_{op t,D}(D_m)$ approaches a constant value for $W_m^+>$25 (see Fig. 1 at page 259 in QR), (ii) the explored range of wall velocities has been $W_m^+<$16 in the previous experimental works and $W_m^+<$18 in the previous numerical works (except for a few cases at $W_m^+$=27 in QR), (iii) QR have found that the drag reduction does not change significantly with $W_m$ for $W_m^+>$20, which greatly simplifies the analysis and it implies that an analysis through $S$ is not needed at these high wall velocities, (iv) it will be shown in Section~\ref{sec:minimal} that the net energy saving may be positive only in the range $W_m^+<$7.

\begin{figure}
\psfrag{D}{$D_m^+$}
\psfrag{T}{$T^+_{opt,D}$}
\centering
\includegraphics[width=\textwidth]{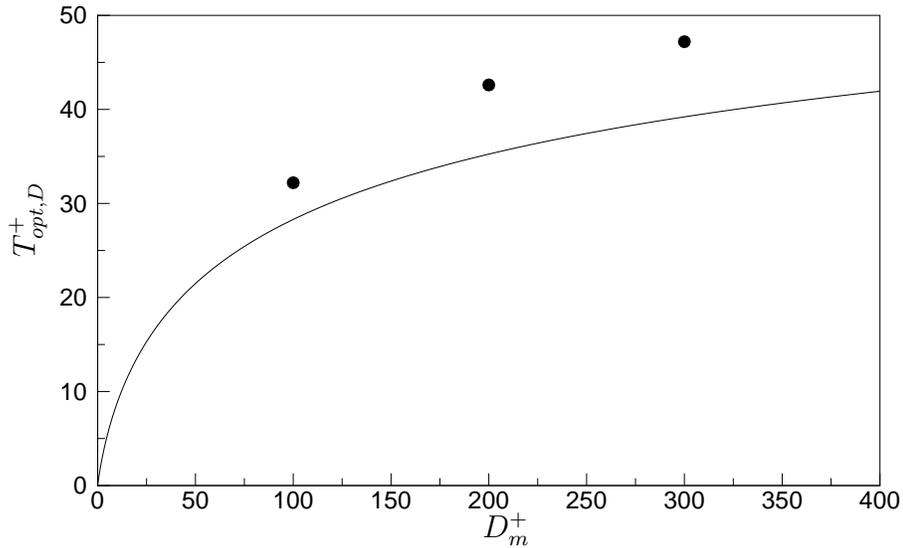}
\caption{Optimum period of oscillation $T^+_{opt,D}$ as function of $D_m^+$. The solid line represents equation (\ref{eq:Topt-Dm}), while dots indicate the values of $T_{opt,D}^+$ determined from the DNS data.}
\label{fig:figure_5}
\end{figure}

The fact that the prediction for $T_{op t,D}$ is good over most of the explored range is further confirmed by Fig. \ref{fig:figure_5}, which shows that the optima computed by the DNS data compare satisfactorily with $T_{opt,D}=T_{opt,D}(D_m)$ obtained by (\ref{eq:Topt-Dm}).

\section{Minimal oscillating conditions and estimates of $DR$(\%) and $P_{net}$(\%) as functions of $W_m^+, D_m^+, T^+$}
\label{sec:minimal}

\begin{figure}
\psfrag{T}{$T^+$}
\psfrag{W}{$W_{m}^+$}
\psfrag{D}{$D_{m}^+$}
\centering
\includegraphics[width=\textwidth]{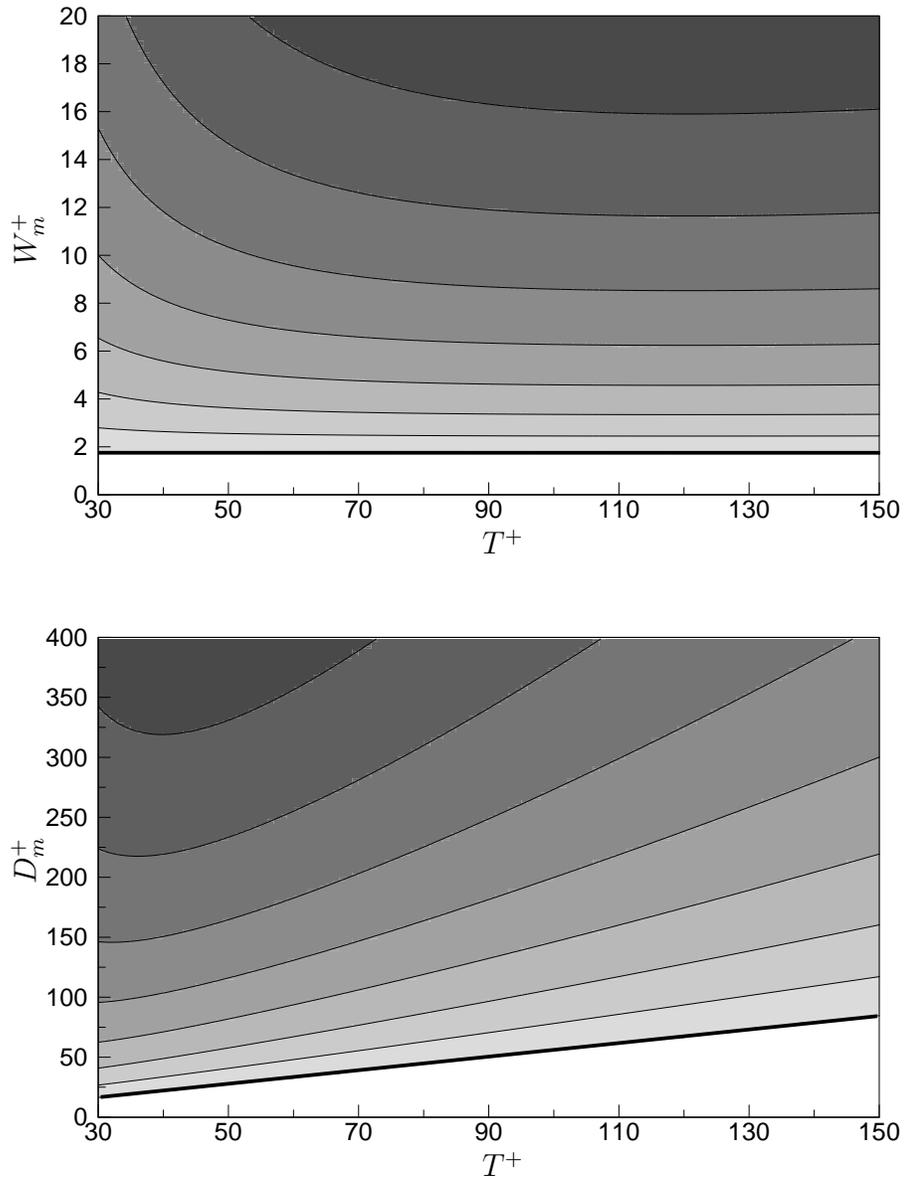}
\caption{Contours of the amount of drag reduction as function of $W_m^+$, $T^+$ (top) and $D_m^+$, $T^+$ (bottom). Darker colours indicate higher drag reduction and contour increments are from zero (thick lines) by 5\%.}
\label{fig:figure_6}
\end{figure}

The idea that a finite intensity of the forcing is needed to affect the turbulent friction is contained in the definition of $S$, where a threshold velocity $W_{th}$ is introduced. The fact that the regression line in Fig. \ref{fig:figure_3}
crosses the abscissa at $S=S_{min}=0.0063>0$ implies that the wall must oscillate with a minimal velocity $W_{m,min}>W_{th}$ (or with a minimal displacement $D_{m,min}>W_{th}T/\pi$) to achieve drag reduction. From (\ref{eq:S}) it follows that
\begin{equation}
\label{eq:minimal}
\left\{
\begin{array}{l l}
W_{m,min}^+ \\
D_{m,min}^+ \\
\end{array}
\right\}
=
\left\{
\begin{array}{l l}
1 \\
T^+/\pi \\
\end{array}
\right\}
 W_{th}^+
\exp \left[ \frac{S_{min}}{2} \sqrt{\frac{T^+}{\pi}} \exp \left( \ell_a^+ \sqrt{\frac{\pi}{T^+}} \right) \right].
\end{equation}
These minimal quantities are displayed in the contours plots of Fig. \ref{fig:figure_6} as the zero drag reduction curves. The drag reduction for a $(T^+,W_m^+)$ pair, computed by (\ref{eq:S}) and (\ref{eq:fit}), is represented in Fig. \ref{fig:figure_6} (top graph) for $30 \leq T^+ \leq 150$. $W_{m,min}$ grows unbounded as $T$ decreases because a stronger wall forcing is needed to affect the turbulent flow as the penetration depth $\delta_s \sim \sqrt{T}$ of the Stokes layer vanishes. $W_{m,min}^+$ becomes approximately constant at higher periods of oscillations, say $T^+>30$, and its value $\approx 1.8$ is of the order of magnitude of the near-wall spanwise velocity fluctuations. It also follows that $D_{m,min}^+ \approx D_1 T^+$, $D_1=W_{m,min}^+(T^+=150)/\pi=0.57$, for $30 \leq T^+ \leq 150$.

Fig. \ref{fig:figure_6} (bottom graph) shows the drag reduction corresponding to a  $(T^+,D_m^+)$ pair.
$D_{m,min}^+(T_{opt,W}^+=125)\approx 70$ compares well with the spanwise width of a low-speed streak, $\approx 20-60 \nu/u_\tau$ from flow visualizations of the near-wall turbulence \citep{hirata-kasagi-1979,ricco-2004}. This result confirms that these structures need to be sufficiently swept laterally to achieve drag reduction \citep{baron-quadrio-1996}. The minimal conditions are in good agreement with the experimental results by Raskob \& Sanderson (personal communication\footnote{B. Raskob and R. Sanderson presented these results at the APS Division for Fluids Dynamics Meeting in San Francisco, California in 1997 with the title {\em Turbulent drag reduction due to an oscillating cross-flow}.}), who first showed that a displacement $D_m^+ \approx 80$ is needed for drag reduction when $80<T^+<1000$.
The existence of the minimal conditions is also confirmed by the analysis of \citet{quadrio-ricco-2003}. They have found that the wall-shear stress is not affected at the beginning of the oscillation, when the wall velocity is small. Figs. 2 and 3 in their paper show that the space-averaged $\tau_x$ changes by less than 1$\%$ when $W^+ < 2$. This value compares well with $W_{m,min}^+$ and it remains essentially unchanged as $T^+$ varies for $50<T^+<200$, similarly to the behaviour of $W_{m,min}^+$ in the same $T^+$ range (see Fig. \ref{fig:figure_6}, top graph).

An analogous behaviour emerges from the experimental study by \cite{mao-hanratty-1986} and by the numerical works by \cite{ismael-cotton-1996} and \cite{cotton-2007}, who studied the variation of wall-shear stress in turbulent pipe flows subjected to streamwise oscillations of the pressure gradient. They found that for small forcing amplitudes the flow responds linearly and the mean value of the velocity gradient at the wall is unaffected.

The expressions (\ref{eq:minimal}) contain the parameter $S_{min}$. Table \ref{tab:Smin} shows that this quantity varies substantially when $W_{th}^+$ and $\ell_a^+$ change by a small amount. However, the minimal conditions remain almost unvaried, as revealed by the values of $W_{m,min}^+$ at $T^+=125$ in Table \ref{tab:Wmin}.

We now turn to the prediction of the net power saving $P_{net}$, defined as the difference between the power saved thanks to the wall motion, i.e. the drag reduction when $U_b$ is constant, and the power $P_{sp}$ spent to move the walls against the viscous stresses. These quantities are expressed as percentage of the power spent to drive the fluid through the unmanipulated channel, $2 \tau_x^+ U_b^+$, where $U_b^+=15.88$ at $Re_\tau=200$. $P_{sp}$ over an interval $t_f - t_i$ is
\begin{equation}
\label{eq:power-spent}
P_{sp}= \frac{L_x L_z}{t_f - t_i} \int_{t_i}^{t_f} \left( \tau^{\ell}_z + \tau^{u}_z \right) W \mathrm{d} t,
\end{equation}
where $\ell$ and $u$ denote the lower and upper walls. Inasmuch as $\tau_z(t)$ is well predicted by the laminar Stokes solution (see Section~\ref{sec:laminar-analysis}), (\ref{eq:power-spent}) becomes

\begin{equation}
\label{eq:power-spent-percent}
P_{sp}(\%)=\frac{100}{\tau_x^+ U_b^+ T^+} \int_0^{T^+} W^+ \frac{\partial w^+_s}{\partial y^+}\Bigg{|}_{y^+=0} \mathrm{d} t =
\frac{100 (W_m^{+})^2}{2 U_b^+}\sqrt{\frac{\pi}{T^+}}=\frac{100 (W_m^{+})^3}{2 U_b^+ Re_\delta}.
\end{equation}

\begin{figure}
\psfrag{T}{$T^+$}
\psfrag{W}{$W_{m}^+$}
\psfrag{D}{$D_{m}^+$}
\centering
\includegraphics[width=\textwidth]{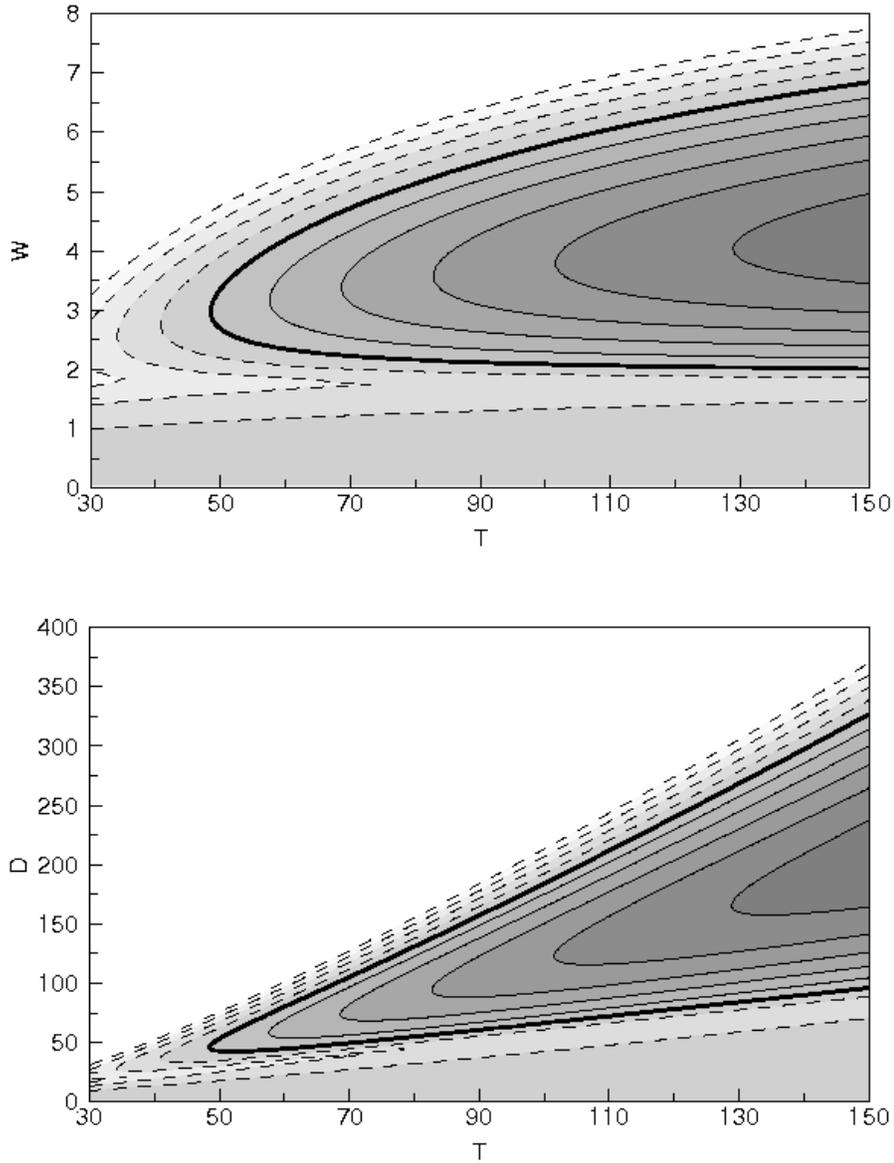}
\caption{Contours of the percent net energy saving $P_{net}$ as function of $W_m^+$, $T^+$ (top) and $D_m^+$, $T^+$ (bottom). Darker colours indicate higher net energy savings, contour increments are by 1\% and dashed lines are for negative values (only values for $P_{net} \geq -4\%$ are shown). The thicker curves denote a null net energy saving.}
\label{fig:figure_7}
\end{figure}

$P_{net}(T^+,W_m^+)$ and $P_{net}(T^+,D_m^+)$ are displayed in Fig. \ref{fig:figure_7}. Steep changes occur when $P_{net}<0$, whereas regions with $P_{net}>0$ present more gradual variations.
When compared with DNS results by QR, the period $T^+ \approx 60$ at $W_m^+=4.5$ needed for $P_{net}>0$ is well captured. For $T^+=150$, $P_{net}>0$ for $2 < W_m^+ < 7$, while in QR this interval is $1.5 < W_m^+ < 9$.

By setting
\[
\frac{\partial P_{net}}{\partial W_m^+}\Bigg{|}_{T^+} = 0, \hspace{0.25 in}
\frac{\partial^2 P_{net}}{\partial W_m^{+2}}\Bigg{|}_{T^+} < 0,
\]
the maximum wall velocity and displacement which give $P_{net,max}$, the maximum $P_{net}$ at fixed $T^+$, are found:
\[
\left\{
\begin{array}{l l}
W_{net,max}^+ \\
D_{net,max}^+ \\
\end{array}
\right\}
=
W_1
\left\{
\begin{array}{l l}
1 \\
T^+/\pi \\
\end{array}
\right\}
\exp \left( - \frac{\ell_a^+}{2} \sqrt{\frac{\pi}{T^+}} \right),
\]
where $W_1=(S_1 U_b^+/50)^{1/2}=6.55$ for $Re_\tau=200$. It follows that
\[
P_{net,max}(T^+)=S_1 \sqrt{\frac{\pi}{T^+}}  \exp \left( - \ell_a^+ \sqrt{\frac{\pi}{T^+}} \right)\left(P_1 - \ell_a^+ \sqrt{\frac{\pi}{T^+}}\right) + S_2,
\]
where $P_1 = 2\ln(W_1/W_{th}^+) - 1 = 1.7$ for $Re_\tau=200$. The location of the overall maximum $P_{net}$ is well estimated at $T^+=150$ and $W_m^+=4.2$, while its value is slightly underpredicted: 5.6\% instead of 7.3\% (QR).

\section{Effect of Reynolds number}
\label{sec:Reynolds}

\begin{figure}
\psfrag{T}{$T^+$}
\psfrag{D}{$DR(\%)$}
\centering
\includegraphics[width=\textwidth]{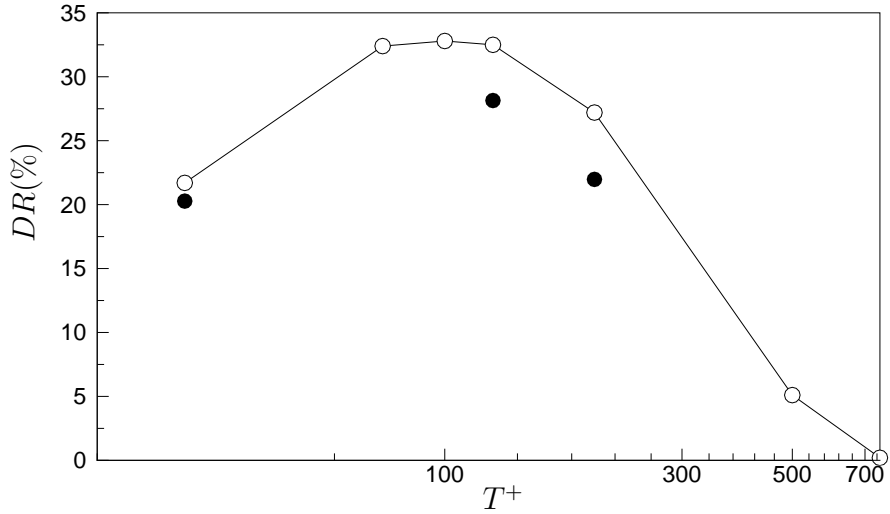}
\caption{Drag reduction as function of $T^+$ for $W_m^+=12$ at $Re_\tau=200$ ($\circ$, data from QR) and $Re_\tau=400$ ($\bullet$).}
\label{fig:figure_8}
\end{figure}

Our calculations show that slightly lower amounts of drag reduction are obtained by doubling the value of the Reynolds number from $Re_\tau=200$ to $Re_\tau=400$ for $W_m^+=12$ and $T^+=30$, 125 and 200, as shown in Fig. \ref{fig:figure_8}. The wall-shear stress reduction decreases from 21.7\% to 20.3\% at $T^+=30$ (6.6\% change), from 32.5\% to 28.1\% at $T^+=125$ (13.4\% change), and from 27.2\% to 22\% at $T^+=200$ (19.2\% change). These variations thus increase with $T$ for fixed $W_m$. This result is in broad agreement with the analysis by \citet{choi-xu-sung-2002}, although a smaller computational domain and a lower spatial resolution were employed in their analysis. Their values of drag reduction at $Re_\tau=200$ are lower than ours and their drag reduction variations with $Re_\tau$ at $W_m^+=10$ are more significant, namely 11.2\% at $T^+=50$, 25.1\% at $T^+=100$, 29.9\% at $T^+=150$, and 29.3\% at $T^+=200$.

Previous experimental studies indicate that the drag reduction does not vary with the Reynolds number for (low) values of this parameter. For example, \citet{ricco-wu-2004} have found that in a turbulent boundary layer the drag reduction does not change as the Reynolds number based on the momentum thickness $\theta$ and the free-stream velocity $U_\infty$ increases from $Re_\theta$=500 to 1400 (which correspond\footnote{The Reynolds number $Re_\theta$ for a free-stream boundary layer can be converted to $Re_\tau=\delta u_\tau/\nu$ (where the boundary layer thickness $\delta$ is the wall-normal distance where the mean velocity is 0.99$U_\infty$) for easiness of comparison with the channel and pipe flow data. By assuming that $\delta = 10 \theta$ \citep{bogard-thole-wall-1998} and by using $C_f = 2(u_\tau/U_\infty)^2 = 0.025 Re_\theta^{-0.25}$ \citep{kays-crawford-1993}, one
arrives at:
\begin{equation}
\label{eq:re-tau-theta}
Re_\tau=1.118 Re_\theta^{0.875}, Re_\theta < 3000.
\end{equation}
}
to $Re_\tau$=257 and $Re_\tau$=633), when $D_m^+ \approx 240$.
\citet{choi-graham-1998} have shown that in a pipe flow modified by circular wall oscillations the wall-shear stress reduction remains within the range of the experimental uncertainty when the Reynolds number $Re$ based on the bulk velocity and the pipe diameter increases from $Re$=23,300 to 36,300 (which correspond\footnote{The Reynolds number $Re$ for pipe flow can be converted to $Re_\tau$ based on $u_\tau$ and pipe radius by using the following formula given in \citet{pope-2000} at pages 292-293:
\begin{equation}
\label{eq:re-conv}
Re=4 \sqrt{2} Re_\tau \left[ 2 \log_{10}\left( 4 \sqrt{2} Re_\tau \right) - 0.8\right].
\end{equation}
} to $Re_\tau=650$ and $Re_\tau=962$).
The uncertainty range on the {\em percent} drag reduction was about 10\% for \citet{ricco-wu-2004} (see Table 1 at page 45 in their paper), and about 20\% for \citet{choi-graham-1998} (see Fig. 2 at page 8 in their paper). It was higher in the second case probably because the wall-shear stress has been determined via measurements of the mass-flow rate and pressure drop, while \citet{ricco-wu-2004} have measured directly the mean streamwise velocity in the viscous sublayer. It is thus possible that the drag reduction variations have not been detected because of the high experimental uncertainty. This conjecture is further supported by the fact that our numerical uncertainty is lower than the ones by \citet{ricco-wu-2004} and \citet{choi-graham-1998} on account of the difficulty of such experiments and the high accuracy of our computations (refer to section 2.4 at page 256 in QR). We also note that the majority of the experiments by \citet{ricco-wu-2004} have been conducted at relatively low $T^+$ ($T^+ \approx 50$, $D_m^+ \approx 240$, $W_m^+ \approx 15$ - see Fig. 13 in their paper), where the change of the drag reduction with the Reynolds number is weak, as shown in Fig. \ref{fig:figure_8}.

Our results might also explain why numerical works have revealed higher drag reduction values than the experimental ones (see Fig. 14 at page 51 in Ricco and Wu, 2004), the latter generally conducted at higher Reynolds numbers. The effect of Reynolds number however warrants further investigation as we have only explored a portion of the $(T,W_m)$ space. It would be of interest to verify the existence of a {\em maximum} Reynolds number above which the wall-shear stress is unchanged, and to investigate how it varies with $T$ and $W_m$.

The effect of the Reynolds number on the maximum net energy saving can be estimated by considering separately the changes on the power spent $P_{sp}$ to move the walls and on the power saved through the wall motion, i.e. the drag reduction. As shown by (\ref{eq:power-spent-percent}), $P_{sp}(\%)$ varies with the Reynolds number only because of a change of $U_b^+$, which can be expressed as follows:
\begin{equation}
\label{eq:Ub}
U_b^+ = \frac{Re}{2 Re_\tau} = 7.715 Re_\tau^{0.136},
\end{equation}
where we have used $Re=15.43 Re_\tau^{1.136}$ given by \cite{pope-2000} at page 279. ($U_b^+$ at $Re_\tau=200, 400$ obtained by our calculations is less than $1$\% different from the one given by (\ref{eq:Ub})). It follows that
\begin{equation}
\label{eq:Psp}
P_{sp}(\%) = 6.481 (W_m^+)^2 \sqrt{\frac{\pi}{T^+}} Re_\tau^{-0.136}.
\end{equation}
The power spent for $W_m^+=4.5$ and $T^+=125$, namely for the oscillating conditions at which the maximum net energy saving occurs (see Section \ref{sec:minimal}), decreases from $P_{sp}$=10.1\% at $Re_\tau=200$ to $P_{sp}$=9.2\% at $Re_\tau=400$. By assuming that the drag reduction for these oscillation conditions decreases by the same amount of the drag reduction at $W_m^+=12$ and $T^+=125$, the maximum net energy saving is estimated to decrease slightly from $P_{net,max}$=7.1\% for $Re_\tau=200$ to 5.7\% for $Re_\tau=400$. We have assumed that $P_{net,max}$ does not change location in the $(T^+,W_m^+)$ space, which remains to be verified. Such change would only be due to a shift of the drag reduction peak, i.e. $T^+_{opt,W}$, and not to $P_{sp}$, which only changes in magnitude when $Re_\tau$ varies, as indicated by (\ref{eq:Psp}). The data in Fig. \ref{fig:figure_8} suggest that $T_{opt,W}^+$ should not vary much as $Re_\tau$ increases.

\section{Analysis of available datasets}
\label{sec:validation-datasets}

Many drag reduction data, either from DNS or experiments, have been reported in the literature for the oscillating wall technique and they are all affected by various sources of error. Accuracy concerns for DNS-based datasets arise mostly from issues related to the spatio-temporal discretization and the computational procedures. In laboratory experiments hot-wire anemometry has often been employed and the measurements may have been biased by errors due to the highly three-dimensional flow field. A comparison among all the drag reduction data have been carried out by QR in the space of the parameters $(T,W_m)$, but the analysis did not convey the desired information about the accuracy because of the high scatter of the data. We attempt here to attain a clearer picture of the state of the art by comparing the $S$-based prediction of drag reduction with the wall-shear stress measured in the laboratory or computed via DNS.

\begin{figure}
\psfrag{S}{$S$}
\psfrag{D}{$DR (\%)$}
\centering
\includegraphics[width=\textwidth]{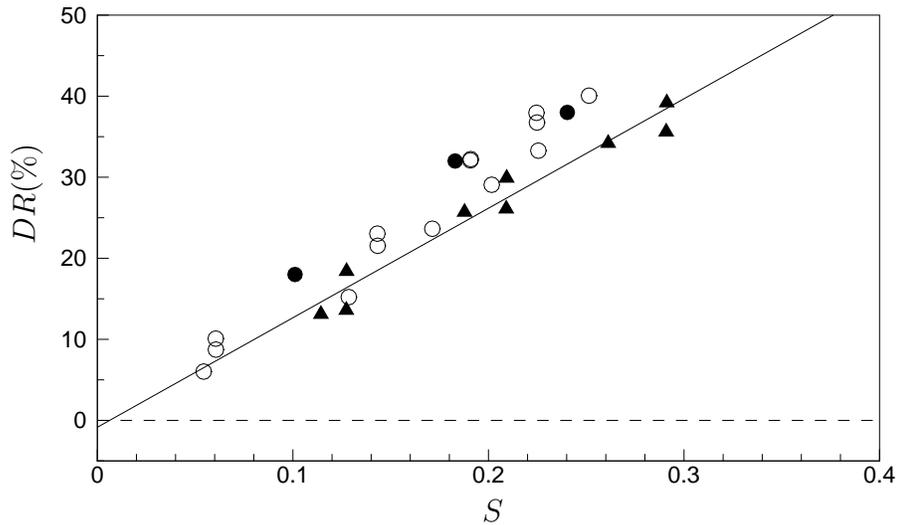}
\caption{Drag reduction data as function of $S$ from three numerical studies at $Re_\tau=200$. Results from \citet{baron-quadrio-1996} ($\bullet$) and \citet{choi-xu-sung-2002} (channel flow) ($\blacktriangle$), \citet{quadrio-sibilla-2000} (pipe flow) ($\circ$).}
\label{fig:figure_9}
\end{figure}

Fig. \ref{fig:figure_9} shows a first group of DNS results. The three datasets by \citet{baron-quadrio-1996} and \citet{choi-xu-sung-2002} for a channel flow and by  \citet{quadrio-sibilla-2000} for a pipe flow are all computed at $Re_\tau=200$ and all broadly agree with one another. Indeed, the numerical accuracy is similar in the three cases. The disparity between the data and the prediction is of the same order of their accuracy (which is lower than the one of the present analysis) in terms of the percent drag reduction. The best correlated dataset is the one by \citet{choi-xu-sung-2002}.

\begin{figure}
\psfrag{S}{$S$}
\psfrag{D}{$DR (\%)$}
\centering
\includegraphics[width=\textwidth]{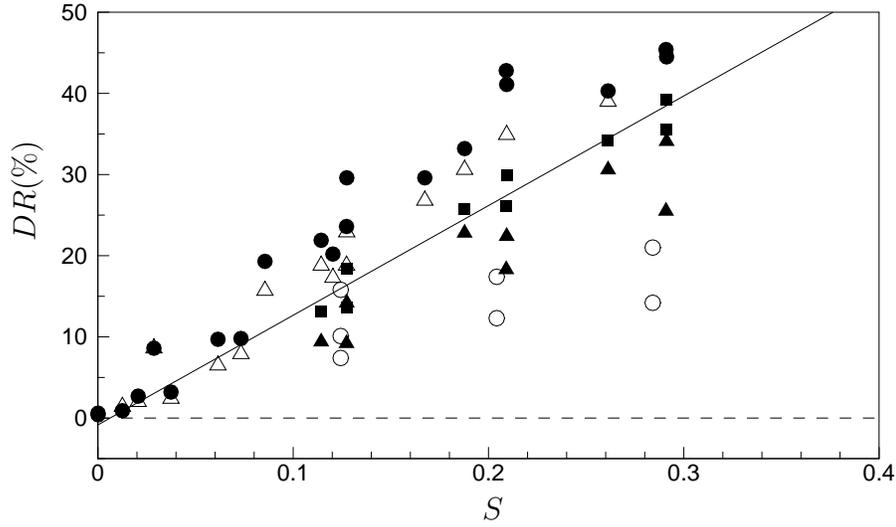}
\caption{Drag reduction data as function of $S$ from \citet{choi-xu-sung-2002}: channel flow at $Re_\tau=100$ ($\bullet$), $Re_\tau=200$ ($\blacksquare$), $Re_\tau=400$ ($\blacktriangle$), pipe flow at $Re_\tau=150$ ($\triangle$), channel and pipe flow with $T^+ =200$ ($\circ$).}
\label{fig:figure_10}
\end{figure}

\citet{choi-xu-sung-2002}'s data in Fig. \ref{fig:figure_10} show higher drag reductions for lower $Re_\tau$. Almost all the data points at $Re_\tau=150$ for the pipe flow fall between the channel flow data at $Re_\tau=100$ and $Re_\tau=200$, which may indicate a negligible influence of the flow geometry. All the data with $T^+=200$ are below the linear trend, confirming that the $S$-scaling occurs only for $T^+<150$ (QR).

\begin{figure}
\psfrag{S}{$S$}
\psfrag{D}{$DR (\%)$}
\centering
\includegraphics[width=\textwidth]{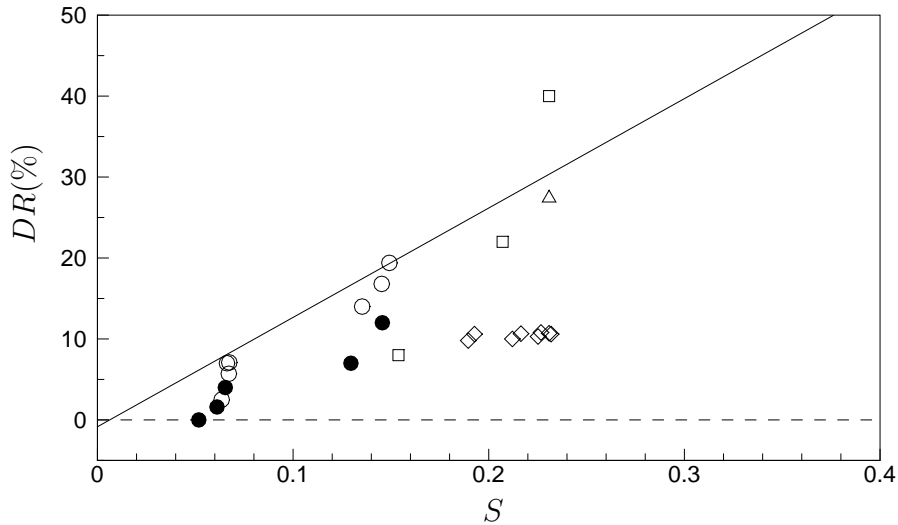}
\caption{Drag reduction data as function of $S$ from various numerical studies. Results from \citet{jung-mangiavacchi-akhavan-1992}: channel flow at $Re_\tau=200$ ($\square$), \citet{nikitin-2000}: pipe flow at $Re_\tau=147$ for $T^+>150$ ($\bullet$) and $T^+<150$  ($\circ$), \citet{dhanak-si-1999} (boundary layer flow) ($\diamond$), and \citet{miyake-etal-1997}: channel flow at $Re_\tau=150$ ($\triangle$).}
\label{fig:figure_11}
\end{figure}

In Fig. \ref{fig:figure_11}, the DNS studies by \citet{jung-mangiavacchi-akhavan-1992} for a channel flow at $Re_\tau=200$ and by \citet{nikitin-2000} for pipe flow at $Re_\tau=147$ ($Re=4000$, see formula (\ref{eq:re-conv})) employed the lowest spatial resolution and the shortest integration time. The comparison with the $S$-estimates is not satisfactory for \citet{jung-mangiavacchi-akhavan-1992}'s data, while \citet{nikitin-2000}'s points fall slightly below the straight line. \citet{jung-mangiavacchi-akhavan-1992} show a 10$\%$ increase of wall friction for $T^+ \approx 500$, $W_m^+ \approx 13$, which has not been confirmed by other works. \citet{dhanak-si-1999}'s data do not follow the expected trend probably because their analysis is based on a simplified  model and not on the full Navier--Stokes equations. However, their analysis was not aimed at an accurate calculation of the wall-shear stress, but at improving the physical understanding. Despite the model limitations, they predicted $T^+_{opt,W} \approx 90$. \citet{miyake-etal-1997}'s point at $Re_\tau=150$ is slightly lower than expected. They might have computed a higher drag reduction (and thus attain a better agreement with our prediction), had they continued the simulation for a longer time (see Fig. 1 at page 203 of their paper, where the wall friction appears likely to decrease further).

\begin{figure}
\psfrag{S}{$S$}
\psfrag{D}{$DR (\%)$}
\centering
\includegraphics[width=\textwidth]{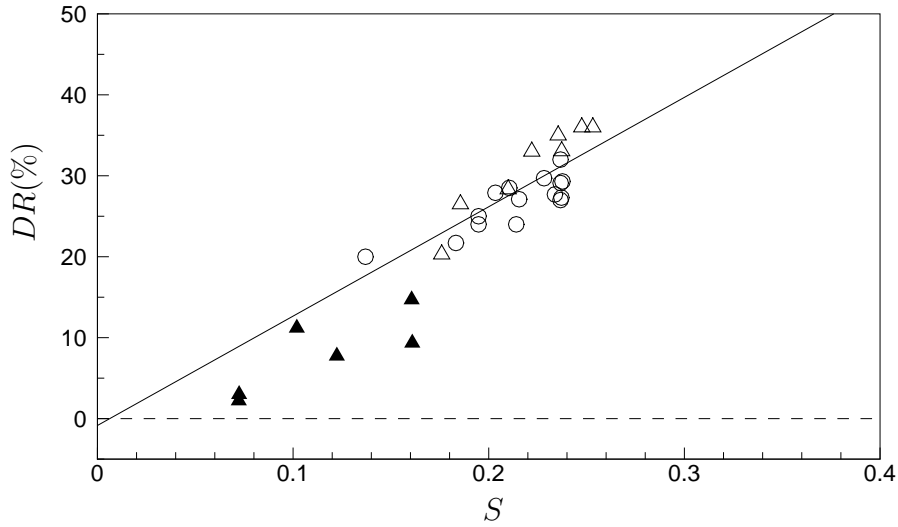}
\caption{Drag reduction data as function of $S$ from two experimental studies on free-stream boundary layer flows. Results from \citet{ricco-wu-2004} ($\circ$) at $Re_\tau=257, 451, 633$ for $T^+<150$ and \citet{skandaji-1997} for $T^+<150$ ($Re_\tau=375, 463$) ($\triangle$) and $T^+>150$ ($Re_\tau=711$) ($\blacktriangle$).}
\label{fig:figure_12}
\end{figure}

Fig. \ref{fig:figure_12} shows that the best correlated experimental dataset is the one by \citet{skandaji-1997} (some of these data are published in \citep{laadhari-skandaji-morel-1994}), for a free-stream boundary layer at $Re_\theta=770$ ($Re_\tau=375$ when formula (\ref{eq:re-tau-theta}) is used), $Re_\theta=980$ ($Re_\tau=463$) and $Re_\theta=1600$ ($Re_\tau=711$). The points at $Re_\tau=711$ are at $T^+ > 150$ and show the discrepancy due to the wall motion uncoupling from the turbulence dynamics. The points with $T^+ < 150$ (actually $T^+ \leq 100$) show a good correlation with the straight line. This is expected as the $Re_\tau$ effect is comparable with the experimental uncertainty at low $T^+$.
The experimental data for a free-stream boundary layer by \citet{ricco-wu-2004} at $Re_\theta=500$ ($Re_\tau=257$), $Re_\theta=950$ ($Re_\tau=451$) and $Re_\theta=1400$ ($Re_\tau=633$) agree with our prediction. The Reynolds number effect is small as most of the data are for $30 < T^+ < 70$.

\begin{figure}
\psfrag{S}{$S$}
\psfrag{D}{$DR (\%)$}
\centering
\includegraphics[width=\textwidth]{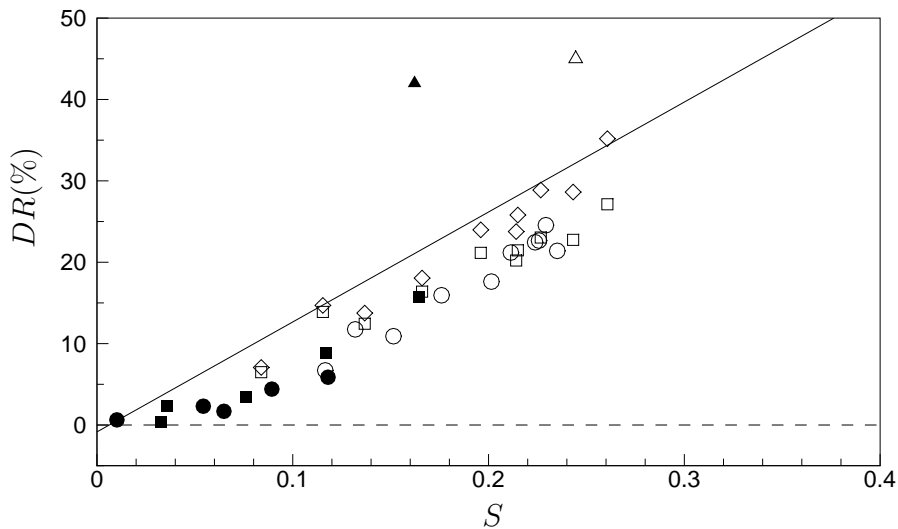}
\caption{Drag reduction data as function of $S$ from three experimental studies. Results from \citet{trujillo-bogard-ball-1997} (boundary layer flow) at $Re_\tau=633$ for $T^+<150$ (biased $\square$ and corrected values $\diamond$), $T^+>150$ ($\blacksquare$). Results from \citet{choi-graham-1998} (pipe flow) at $Re_\tau=650, 962$ for $T^+<150$ ($\circ$) and $T^+>150$ ($\bullet$), and from \citet{choi-debisschop-clayton-1998} (boundary layer flow) at $Re_\tau=549$ for $T^+<150$ ($\triangle$) and $T^+>150$ ($\blacktriangle$).}
\label{fig:figure_13}
\end{figure}

\citet{trujillo-bogard-ball-1997}'s experimental data for a free-stream boundary layer at $Re_\theta=1400$ ($Re_\tau=633$), presented in Fig. \ref{fig:figure_13}, show the correct slope. They are however lower than the linear fit, probably because of the bias caused by the spanwise component of velocity on the hot-film measurements in the proximity of the wall (Trujillo, personal communication; Choi and  Clayton 2001; Ricco and Wu, 2004). The same data have been corrected in \citet{trujillo-1999}, thus reaching a better agreement with the line (only the corrected data with $T^+ < 150$ are included for clarity in Fig. \ref{fig:figure_13}). The fact that the data at low $T$ by \citet{skandaji-1997}, \citet{ricco-wu-2004} and \citet{trujillo-1999} fall near the linear regression could be a sign that the two flat-plate geometries, i.e. a free-stream boundary layer as in the experiments and a pressure-driven channel flow as in our simulations, might have the same or very similar drag reduction properties. The pipe flow data by \citet{choi-graham-1998} at $Re_\tau=650$ and $Re_\tau=962$ also have the same slope, but are lower than the linear regression, probably on account of the high values of $Re_\tau$. The two points by \citet{choi-debisschop-clayton-1998} for boundary layer flow at $Re_\tau=549$ are higher than expected.

\section{Summary}
\label{sec:conclusions}

We have presented a study of the drag reduction effects of spanwise wall oscillations on a turbulent channel flow based on the direct numerical simulation of the incompressible Navier--Stokes equations. A scaling parameter $S$, first proposed by \citet{choi-xu-sung-2002} and later studied by \citet{quadrio-ricco-2004b}, is further considered and used to improve our understanding of the main properties of this drag reduction technique.

We have discovered that $S$ is an excellent predictive tool for drag reduction for $W_m^+ \leq 40$ and $30 \leq T^+ \leq 150$. This region of the parameters space includes most of the published drag reduction data and is of most practical importance. The existence of two optimal drag reduction periods, $T_{opt,D}$ and $T_{opt,W}$, deduced from the expression for $S$ and verified through our DNS database, is now clearly established. $T_{opt,D}$, the period of oscillation which guarantees the maximum drag reduction at fixed peak-to-peak wall displacement $D_m$, primarily concerns experimentalists, who are forced to vary $T$ with $D_m$ constant. $T_{opt,W}$, the optimum period of oscillation at fixed maximum wall velocity, is the quantity typically searched for by numerical investigators. The function $T_{opt,D}=T_{opt,D}(D_m)$ has been determined from the expression for $S$ and confirmed via numerical experiments carried out to the purpose. We have also established the minimal wall forcing conditions leading to drag reduction and obtained plots of the drag reduction and of the net energy saving as functions of the oscillation parameters. Despite some discrepancy with the available data, such charts might prove useful for the design of a drag-reducing device. Further work is necessary to establish how the wall-shear stress changes from the unperturbed condition in the limits $T^+ \rightarrow 0, \infty$ for $D_m^+$ or $W_m^+$ constant.

The amount of drag reduction has been found to decrease slightly as the Reynolds number varies from $Re_\tau=200$ to $Re_\tau=400$ when $W_m^+=12$. This effect amplifies as the period of oscillation increases. Further study should be conducted to investigate how the optimal periods of oscillation change with the Reynolds number. We have also estimated that the maximum net energy saving decreases slightly with $Re_\tau$. The existing numerical and experimental data have been re-examined through a comparison between the measured amounts of drag reduction and their $S$-based estimates: differences have been discussed on a case-by-case basis in relation to the accuracy of each dataset.

\section*{Acknowledgments}

The extensive use of the computing system, run by Professor Paolo Luchini at the University of Salerno, is gratefully acknowledged. We would like to thank Dr Bill Raskob for providing us with his experimental results and the referees for the helpful comments.

\section*{Nomenclature}
\begin{tabbing}

\bf{Symbol}\qquad\= \quad\= \bf{Definition}\quad\= \\ \\

$a_m$		\> \>	maximum acceleration of Stokes layer, $\rm m/s^2$ \\
$C_S$		\> \>	correlation coefficient in $S$-scaling regression analysis, dimensionless \\
$C_f$		\> \>	skin-friction coefficient $C_f = 2 \tau_x / (\rho U_b^2)$ for channel flows
\\ and $C_f = 2 \tau_x / (\rho U_\infty^2)$ for boundary layer flows, dimensionless \\
$D_1$		\> \>	coefficient used for prediction of $D_{m,min}$, dimensionless \\
$D_m$		\> \>	peak-to-peak maximum wall displacement, $\rm m$ \\
$D_{m,min}$	\> \>	minimum value of $D_m$ for drag reduction, $\rm m$ \\
$DR$		\> \>	percent drag reduction, (\%) \\
$h$		\> \>	half channel height or pipe radius, $\rm m$ \\
$l$		\> \>	superscript indicating lower wall \\
$l_a$		\> \>	wall-normal distance at which $a_m$ is computed, $\rm m$ \\
$l_m$		\> \>	wall-normal distance used in the computation of $S$, $\rm m$ \\
$L_x$		\> \>	streamwise length of computational box, $\rm m$ \\
$L_z$		\> \>	spanwise width of computational box, $\rm m$ \\
$P_{net}$	\> \>	net power saved to drive fluid through channel \\ 
thanks to wall oscillation, $\rm kg$ $\rm m^4/s^3$ \\
$P_{net,max}$	\> \>	maximum $P_{net}$ at fixed $T^+$ at given $Re_\tau$, $\rm kg$ $\rm m^4/s^3$ \\
$P_{sp}$	\> \>	power spent to move walls, $\rm kg$ $\rm m^4/s^3$ \\
$R_{vw}$	\> \>	$R_{vw}=\partial \overline{v' w'}^+/\partial y^+$, $\rm m/s^2$ \\
$Re$		\> \>	Reynolds number based on $U_b$ and $2h$ for channel flows
\\ and on $U_b$ and pipe diameter for pipe flows, dimensionless \\
$Re_\delta$	\> \>	Reynolds number for Stokes layer ($Re_\delta = W_m \delta_s/\nu$), dimensionless \\
$Re_\theta$	\> \>	Reynolds number based on $U_\infty$ and $\theta$, dimensionless \\
$Re_\tau$	\> \>	Reynolds number based on $u_\tau$ and $h$, dimensionless \\
$S$		\> \>	drag reduction scaling parameter, dimensionless \\
$S_1,S_2$	\> \>	constants in equation (\ref{eq:fit}), dimensionless \\
$S_{min}$	\> \>	minimum value of $S$ for drag reduction, dimensionless \\
$t$		\> \>	time, $\rm s$ \\
$t_f$		\> \>	time at which computation of wall-shear stress is completed, $\rm s$ \\
$t_i$		\> \>	time at which computation of wall-shear stress is initiated, $\rm s$ \\
$T$		\> \>	period of wall oscillation, $\rm s$ \\
$T_{opt,D}$	\> \>	optimum period of wall oscillation at fixed $D_m$, $\rm s$ \\
$T_{opt,W}$	\> \>	optimum period of wall oscillation at fixed $W_m$, $\rm s$ \\
$u$		\> \>	superscript indicating upper wall \\
$u_\tau$	\> \>	friction velocity for fixed wall configuration
$\left(u_\tau=\sqrt{\tau_x/\rho}\right)$, $\rm m/s$ \\
$U_b$		\> \>	bulk velocity in channel and pipe flows, $\rm m/s$ \\
$U_\infty$	\> \>	mean free-stream velocity for boundary layer flows, $\rm m/s$ \\
$v'$		\> \>	wall-normal turbulent velocity fluctuating about $x-z$ average, $\rm m/s$ \\
$w$		\> \>	spanwise turbulent velocity, $\rm m/s$ \\
$w'$		\> \>	spanwise turbulent velocity fluctuating about $x-z$ average, $\rm m/s$ \\
$w_s$		\> \>	velocity of Stokes layer, $\rm m/s$ \\
$W$		\> \>	wall velocity, $\rm m/s$ \\
$W_m$		\> \>	maximum wall velocity, $\rm m/s$ \\
$W_{m,min}$	\> \>	minimum value of $W_m$ for drag reduction, $\rm m/s$ \\
$W_{th}$	\> \>	threshold spanwise velocity used in the computation of $S$, $\rm m/s$ \\
$y$		\> \>	vertical direction, $\rm m$ \\
$\delta$	\> \>	boundary layer thickness, i.e. wall-normal distance at which
\\ the mean streamwise velocity equals 0.99$U_\infty$ for boundary layer flows, $\rm m$ \\
$\delta_s$	\> \>	Stokes layer thickness $\left(\delta_s = \sqrt{\nu T/\pi}\right)$, $\rm m$ \\
$\theta$	\> \>	momentum thickness for free-stream boundary layers, $\rm m$ \\
$\nu$		\> \>	kinematic viscosity of the fluid, $\rm m^2/s$ \\
$\rho$		\> \>	density of the fluid, $\rm kg/m^3$ \\
$\tau_x$	\> \>	time- and/or space-averaged streamwise wall-shear stress, $\rm kg$ $\rm m/s^2$ \\
$\tau_z$	\> \>	space-averaged spanwise wall-shear stress, $\rm kg$ $\rm m/s^2$ \\ 
$+$		\> \>	indicates quantities scaled by inner variables, i.e. $u_\tau$ and $\nu$ \\

\end{tabbing}

\bibliographystyle{elsart-harv}

\pagebreak


\pagebreak

\begin{table}
\centering
\begin{tabular}{c|ccc}
$\ell_a^+$ & 6.1 & 6.2  & 6.3  \\
\hline
$W_{th}^+$ & & &  \\
1.5 & 0.0158  &  0.0144       & 0.0130    \\
1.6 & 0.0108  &  0.0097       & 0.0085    \\
1.7 & 0.0061  &  {\bf 0.0063} & 0.0042    \\
1.8 & 0.0014  &  0.0007       & $\sim$0.0 \\
\end{tabular}
\caption{Variation of $S_{min}$ with $\ell_a^+$ and $W_{th}^+$. The value in bold gives the lowest $C_S$.}
\label{tab:Smin}
\end{table}

\newpage

\pagebreak

\begin{table}
\centering
\begin{tabular}{c|ccc}
$\ell_a^+$ & 6.1 & 6.2  & 6.3  \\
\hline
$W_{th}^+$ 	& 	& 		&  	\\
1.5 		& 1.71 	& 1.69 		& 1.68 		\\
1.6 		& 1.75 	& 1.73		& 1.72 		\\
1.7 		& 1.79 	& {\bf 1.79} 	& 1.76 \\
1.8 		& 1.82 	& 1.81		& 1.80 		\\
\end{tabular}
\caption{Variation of $W_{min}^+(T^+=125)$ with $\ell_a^+$ and $W_{th}^+$. The value in bold gives the lowest $C_S$.}
\label{tab:Wmin}
\end{table}

\pagebreak

\pagebreak

\pagebreak

\pagebreak

\pagebreak

\pagebreak

\pagebreak

\pagebreak

\pagebreak

\pagebreak

\pagebreak

\pagebreak

\pagebreak

\newpage

\end{document}